\documentclass{elsart}
\usepackage{graphics}

\begin{document}

\def\etal{{\it{}et~al.}}        
\def\th#1{#1$^{\rm th}$}        
\def\av#1{\langle#1\rangle}     
\def\ra{\rightarrow}            
\def\la{\leftarrow}             
\def\xx{}

\def\psfigure#1#2{\resizebox{#2}{!}{\includegraphics{#1}}}
\def\eref#1{\protect(\ref{#1})}

\textfloatsep=1cm               
\floatsep=1.5cm                 

\begin{frontmatter}
\title{The repton model of gel electrophoresis}
\author[IAS]{G. T. Barkema}
and
\author[SFI]{M. E. J. Newman}
\address[IAS]{Institute for Advanced Study, Olden Lane, Princeton,
NJ 08540.  U.S.A.}
\address[SFI]{Santa Fe Institute, 1399 Hyde Park Road, Santa Fe, NM 87501.
U.S.A.}
\begin{abstract}
  We discuss the repton model of agarose gel electrophoresis of DNA.  We
  review previous results, both analytic and numerical, as well as
  presenting a new numerical algorithm for the efficient simulation of the
  model, and suggesting a new approach to the model's analytic solution.
\end{abstract}
\end{frontmatter}

\section{Introduction}
Gel electrophoresis is a technique of great importance in the
rapidly-growing fields of molecular genetics and genetic engineering.  It
provides a simple and economical way of measuring the lengths of polymer
strands and, if necessary, refining them by length as well.  In this
article we will consider the physics of the agarose gel process, which is
commonly applied to DNA, although it can in principle be applied to other
polymers, such as RNA for example.  (The other common variety of process is
polyacrylamide gel electrophoresis, whose physics is somewhat different and
not well modelled using the techniques described here.)

The basic idea is very simple.  DNA or other polymer strands of a variety
of unknown lengths are deposited in a thin layer of gel composed of agarose
powder and a buffer solution which imparts a charge to each base pair or
monomer.  An electric field is applied horizontally and the charged DNA
migrates under its influence.  The rate of migration is found to depend on
the length of the strands, with shorter strands travelling further in a
given time than longer ones.  After a period of some hours, the initial
mixture of strands will have become separated and spread over several
inches of the gel, allowing the experimenter to measure the relative
concentrations of strands of different lengths, or cut particular areas out
of the gel, thereby isolating strands with a particular range of lengths.

Given the technological importance of the method, there is much interest in
learning precisely what the mechanisms of gel electrophoresis are and how
the migration rate depends on strand length, applied electric field, and
the properties of the agarose gel.  It is known that the agarose forms long
strands in the gel which cross-link and impede the movement of the DNA.
Furthermore, the pores between these strands have a size which is roughly
equal to the persistence length of the DNA, with the result that the DNA is
prevented from moving transversely to its length; its only mode of
transport is a snake-like slithering along its length through the pores of
the gel, a motion which de Gennes~\cite{deGennes71} has dubbed
``reptation'': movement of a polymer along its own length by diffusion of
stored length.  Many attempts have been made to model the dynamics of
reptation, with varying degrees of sophistication, but interestingly one of
the most quantitatively successful of these is also the simplest of the
lot, the so-called ``repton model''.  This model was first introduced by
Rubinstein in 1987~\cite{Rubinstein87} and has been extensively studied by
various authors~\cite{Duke89,Duke90a,Duke90b,LK92,KL93a,KL93b}, including
Widom and co-workers~\cite{WVD91,Widom96,BMW94}, who gave both analytic
solutions of the model for short strand lengths and also extensive
numerical simulations using a number of novel Monte Carlo techniques.

The outline of this paper is as follows.  In Section~\ref{reptonmodel} we
describe the repton model and discuss its relationship to gel
electrophoresis experiments.  In Sections~\ref{analytic}
and~\ref{numerical} we review some of the analytic and numerical results
for the model.  In Section~\ref{secparticles} we introduce a mapping of the
model to a hard-sphere particle model which has allowed us to perform more
accurate numerical calculations and we present preliminary results from these
calculations.  We also discuss briefly how the particle mapping implies a
connection between the repton model and the so-called asymmetric exclusion
models, a link which may ultimately lead to an exact solution of the
problem.  In Section~\ref{conclusions} we give our conclusions.

\section{The repton model}
\label{reptonmodel}
\begin{figure}
\begin{center}
\psfigure{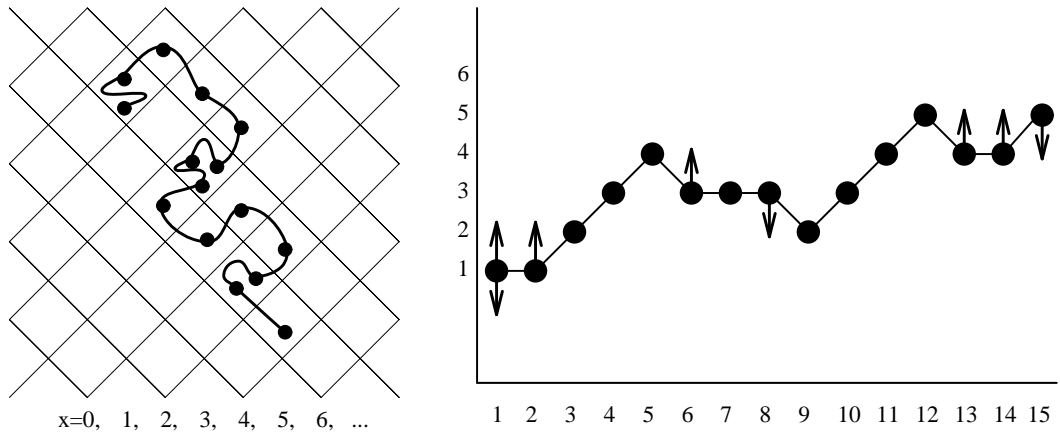}{\textwidth}
\end{center}
\caption{A typical configuration of the repton model (left).  If the reptons
are numbered along the chain, starting with the end in the top-left corner,
we can characterize this configuration by plotting the $x$-position as a
function of the repton number.  This is done in the right side of the
figure.  This is the projected model described in the text.  Allowed moves
are denoted by arrows.}
\label{repton}
\end{figure}

The repton model of polymer reptation in a gel is illustrated in
Figure~\ref{repton}.  The gel is approximated by a collection of pores
arranged on a square lattice.  These are represented by the diagonal grid
of squares in the figure.  (The lattice is shown as two dimensional in
Figure~\ref{repton}, although the real system is three dimensional.
However, as discussed in Section~\ref{subsecproject}, the properties of the
model are independent of the number of dimensions of the lattice, so we may
as well stick with two for clarity.)  We represent the DNA or other polymer
as a chain of $N$ polymer segments or ``reptons'', the black dots in the
figure.  (The reptons are not the same thing as base pairs; rather each one
corresponds to one persistence length of DNA.  Depending on the conditions
of the experiment, the persistence length of DNA is between about 150 and
300~base pairs, as discussed in Section~\ref{subsecvalues}.)  Reptons move
from pore to adjacent pore diagonally according to the following rules:
\begin{enumerate}
\item Reptons in the interior of the chain move only along the line of
  pores occupied by the chain.  Thus a repton in the interior of the chain
  can only move to an adjacent pore if that pore is already occupied by one
  of its neighbouring reptons in the chain.  Note that a consequence of
  this restriction is that if three or more neighbouring reptons in the
  chain find themselves all in one pore, only the two with connections to
  other reptons outside this pore are allowed to move.
\item At least one repton must remain in each pore along the chain but
  otherwise the number of reptons in a pore is unrestricted.  This means
  that a repton in the interior of the chain can only leave a pore if one
  of its neighbours is also in that pore and stays behind when it leaves.
  This second rule gives the polymer some elasticity without making it
  infinitely stretchy.
\item The two reptons at the ends of the polymer chain can move to adjacent
  pores provided that rule~(ii) is not violated.
\end{enumerate}
In a real polymer there are additional excluded-volume effects due to
the finite space occupied by the polymer, as well as a number of other
physical processes which are not included in the repton model.  It is
assumed that these effects make only a small contribution to the
behaviour of the polymer under electrophoresis.

Since we are interested in the rate at which DNA drifts through the gel, we
also need to define the time-scale on which moves take place.  To do this,
we make the assumption that thermal fluctuations continually drive all the
reptons to attempt moves to adjacent squares on the grid.  We assume that
every such move is as likely to be attempted as every other at any
particular time, and we choose the time-scale such that each such move will
be attempted once, on average, per unit time.  (Not all of these moves will
actually take place, since many of them will be rejected for violating one
of the rules above.  However, the moves which {\em are\/} allowed each take
place with equal probability per unit time.  There are no energies
associated with the different states, and so no Boltzmann weights
making one move more likely than another.  The dynamics of the model is
purely entropic.)

The model described so far is just a model of polymer diffusion.  All the
motions are random thermal ones and there is no applied electric field.  To
make a model of electrophoresis, we assign to each repton the same negative
electric charge, mimicking the charging effect of the buffer solution in
the experiment.  Then we apply a uniform electric field to the model along
the $x$-axis (the horizontal direction in the figure), breaking the spatial
symmetry.  As a result, instead of unit rates for the allowed moves in the
positive and negative $x$-directions, the rates become $\exp(-E/2)$ in the
positive direction and $\exp(E/2)$ in the negative one, where $E$ is a new
parameter which is proportional to the applied field (see
Section~\ref{subsecvalues}).  The resulting model describes the qualitative
features of DNA electrophoresis surprisingly well, at least for longer
strands of DNA (above about 1kb).  Many details of the real system are
missing, such as mechanical properties of the polymer, effects of
counterions, the inhomogeneity of the gel, and the gel concentration, and
to be truly realistic the model would have to include these features as
well.  However, we can extract a lot of useful information from the simple
model we have already.

\subsection{The projected repton model}
\label{subsecproject}
The drawing in Figure~\ref{repton} is two-dimensional, but real
electrophoresis experiments are three-dimensional.  Does this matter?  As
we mentioned briefly above, it does not, for the following reason.  We are
interested in calculating the drift velocity of the polymer for the case of
a uniform electric field in the negative $x$-direction, which means we are
interested only in the $x$-coordinates of the polymer; the values of the
other coordinates are irrelevant.  Thus, we might as well project the
polymer onto the $x$-axis and ignore the other axes.  We do precisely this
on the right hand side of Figure~\ref{repton}, where we have numbered the
reptons along the length of the chain and then plotted the $x$-coordinate
as a function of repton number.  This one-dimensional projection of the
model proves very convenient both for analytic and numerical calculations,
and we will mainly be discussing this form of the model from here on.

Comparing the states of the projected model with those of the original
higher-dimensional version, we can easily work out how the dynamics looks
after projection.  First, reptons in the projected chain must either be at
the same height as, or one place higher or lower than their neighbours, and
each possible move consists of moving one repton either up or down one
place.  However, just satisfying these constraints is not enough.
Consider, for instance, the repton numbered~7 in Figure~\ref{repton}.  This
repton is not allowed to move either up or down, even though the final
configuration following such a move would be a perfectly legitimate one.
This fact can easily be verified by going back to the higher-dimensional
representation and examining repton~7.  In fact there is essentially only
one type of move allowed in the projected model, in which a repton which
has one of its neighbours on the level above or below it and one on the
same level is allowed to move up or down respectively.  The reader may like
to verify that such moves are indeed the equivalent of reptation in the
higher-dimensional version of the model.  For the particular configuration
shown in Figure~\ref{repton}, the allowed moves for the projected model are
shown with arrows.

This projection of the repton model onto one dimension works just as well
from three dimensions as from two, or indeed from any higher number of
dimensions.  The final rules for the dynamics of the projected model are
just the same in each case, and so we conclude that the properties of the
model are independent of the dimensions of the lattice on which it is
built.

\subsection{Values of the parameters in the model}
\label{subsecvalues}
Before we can investigate the properties of the repton model, we need to
get an idea of the sort of values that the parameters $N$ and $E$ should
take.  The parameter $N$ measures the length of the polymer in multiples of
the persistence length.  Double-stranded DNA is quite stiff, and has a
persistence length of around 150~bp, or about 400~\AA.  It can become
stiffer still as a result of self-repulsion when it acquires an electric
charge.  Since the size of the charge picked up per base pair depends on
the strength of the buffer solution used in the experiment, the actual
persistence length varies from one experiment to another.  In some cases it
can be as high as 800~\AA.

But $N$ also plays another role in the repton model.  As we said above, the
diagonal lattice of squares in Figure~\ref{repton} represents the pores in
the agarose gel, and the lattice spacing represents the typical pore size.
But in this model, the maximum distance from one repton to the next is one
lattice spacing, so $N$ is also effectively the length of the polymer in
multiples of the pore size.  Thus, the repton model will only be an
effective model of DNA reptation if the persistence length and the pore
size are roughly equal.  The pore sizes in agarose gels have not been
determined with great accuracy, but for the typical 1\% agarose gel they
are estimated to be in the range of 1000 to 3000~\AA.  Thus the two length
scales are indeed of the same order of magnitude, and this goes some way
towards explaining the success of the model.  By contrast, the persistence
length of RNA is only one or two bp or about 3 to 5~\AA\ since RNA is
largely single-stranded.  This length scale differs too greatly from the
agarose pore size to allow the two scales to be well represented by the one
parameter in the repton model, and as a result the model does not make good
predictions about RNA electrophoresis experiments.

The dimensionless parameter $E$ is defined so as to be proportional to the
electric field applied to the system.  The quantity $\exp(E)$ is the ratio
between the probability for a repton to move one step down the slope of the
electric potential and the probability for moving one step up it, and so
should be equal to the ratio of the Boltzmann factors for making those two
moves.  This means that
\begin{equation}
E = {\sqrt2 a q E_f\over kT},
\end{equation}
where $a$ is the lattice parameter (i.e.,~the pore size), $q$ is the charge
per repton (i.e.,~per persistence length), and $E_f$ is the applied
electric field.  The numerator here represents the energy needed to move
a repton a distance $\sqrt2 a$ (the $x$-distance between two nearest
neighbours of the same lattice site) against the electric force $qE_f$
acting on it.

The charge $q$ is on the order of one electron charge $e$ per base pair, or
about $150\times1.6\times10^{-19}$~C per repton.  Using typical values for
the other parameters, we find that
\begin{equation}
E \sim 10^{-3} E_f
\end{equation}
when $E_f$ is measured in Vm$^{-1}$.  This figure is bourne out by the
experimental results of Barkema, Caron and Marko~\cite{BCM96}.

\section{Exact calculations}
\label{analytic}
In the limit of infinite length, Pr\"ahofer~\cite{Prahofer94} has shown
that the diffusion constant of the repton model is exactly equal to
$1/(3N^2)$.  However, the model shows strong finite size corrections for
values of $N$ corresponding to the typical lengths of DNA strands used in
experimental work, making it important that we calculate its behaviour for
finite lengths also.  For short chains one approach which has proved
fruitful is the matrix technique employed by Widom and
co-workers~\cite{WVD91,Widom96}.  It is clear that for a strand of any
finite length one can enumerate all possible states of the projected repton
chain.  Each link of the chain can be in one of three states---horizontal,
or sloping in one of two directions.  (There is an overall translation
degree of freedom, vertically in Figure~\ref{repton}, but for the moment we
are only concerned with the states of the chain itself.)
For a chain of $N$ reptons, there are $3^{N-1}$ such states, and we can
write down a transition rate matrix ${\bf T}(E)$, such that the element
$T_{ij}$ is the probability that the chain will be in a state $j$ if it
was in a state $i$ one move earlier.  The off-diagonal elements take
values proportional to $\e^{\pm E/2}$ if the states $i$ and $j$ differ
by an allowed move of exactly one repton, and are zero otherwise.  The
diagonal ones are chosen to make the sum of the elements in each column
equal to unity.

The probability $p_i$ of a particular state appearing in a large ensemble of
identical chains, in the limit $t\to\infty$, is given by the $i^{\rm th}$
element of the eigenvector corresponding to the slowest-decaying eigenmode
of this matrix.  Widom~\etal~\cite{WVD91} showed that these probabilities
are related to the diffusion velocity $v$ of the chain via
\begin{equation}
v = {1\over N} \sum_i (\e^{E/2} r_i - \e^{-E/2} s_i) p_i,
\end{equation}
where $r_i$ and $s_i$ are respectively the number of reptons which can
lawfully move up and down when the chain is in state $i$.  The diffusion
constant $D$ of the polymer strand in the gel is given in terms of this
velocity by the Nernst-Einstein relation:
\begin{equation}
D = \lim_{E\to0} {v\over NE}
\label{nernst}
\end{equation}
for a system in which there is one dimensionless unit of charge on each
repton.  The only obstacle to the exact calculation of the diffusion constant
then is the limit taken in Equation~\eref{nernst}.  Widom~\etal\ circumvented
this problem by noting that in the zero-field case the solution for the
probabilities $p_i$ is simple---all states are equally probable and $p_i =
3^{-(N-1)}$ for all $i$---and perturbing around this solution to find the
small $E$ behaviour:
\begin{equation}
p_i = {1\over3^{N-1}} [1 + \half a_i E + {\rm O}(E^2)].
\label{linear}
\end{equation}
Starting with the transition rate matrix, they showed that the coefficient
$a_i$ is given by the solution of a set of linear equations
\begin{equation}
\sum_{i\leftrightarrow j} (a_i - a_j) = -2 (s_i - r_i),
\label{tm}
\end{equation}
where the notation ${i\leftrightarrow j}$ means that the state $i$ can be
reached from $j$ by one allowed move.  Along with the subsidiary condition
\begin{equation}
\sum_i a_i = 0
\end{equation}
this set of equations determines the values of the coefficients $a_i$ for
all states of the chain.  Using Equation~\eref{linear} we can then show
that the diffusion constant is given in terms of these coefficients by
\begin{equation}
D = {1\over2N^2} [\mbox{$\frac49$} (N+4) - {1\over3^{N-1}}\sum_i (s_i -
r_i) a_i].
\end{equation}

The only catch of course is that there are $3^{N-1}$ equations
in~\eref{tm}, so that solving them involves the diagonalization of a
$3^{N-1}\times3^{N-1}$ matrix and, sparse though that matrix is, the
problem rapidly grows beyond the capabilities of even the most powerful
computers.  Widom~\etal\ computed the diffusion coefficients for systems
with values of $N$ up to $5$.  These results were extended to $N=12$ by
Szleifer and Bisseling (unpublished) using a special purpose computer.
However, given the exponential increase in the magnitude of the diagonalization
problem with increasing $N$, it seems unlikely that this method will be
taken any further.

\section{Simulation results}
\label{numerical}
In order to study chains of more than twelve reptons, we turn to Monte
Carlo calculations.  In their simplest form, these calculations are a
direct implementation of the repton dynamics described in
Section~\ref{reptonmodel}.  In each step of the simulation we choose a
repton at random from the chain and move it either in the direction of the
electric field or against it, with probabilities proportional to $\e^{\pm
  E/2}$.  In order that we attempt the correct number of moves per unit
time, a move should correspond to a time interval of
\begin{equation}
\Delta t = {1\over N(\e^{E/2} + \e^{-E/2})} = {1\over2N\cosh (E/2)}.
\label{moves}
\end{equation}

We also need to make an estimate of the correlation time $\tau$ of the
simulation, so that we know how many Monte Carlo steps we need to run for.
For small electric fields the typical end-to-end distance of the chain is
$\sqrt{N}$, and on average the chain reaches a new independent
configuration every time it moves a length approximately equal to this
distance.  The correlation time is then given by Equation~\eref{nernst}
thus:
\begin{equation}
\tau = {\sqrt{N}\over v} = {\sqrt{N}\over DNE} = {3N^{3/2}\over E}.
\end{equation}
This estimate diverges as $E\to0$.  However, it is clear that the
divergence is truncated at $E=0$ since in that case the mean square
distance travelled by the polymer in any direction as a function of time is
given simply by
\begin{equation}
\langle x^2 \rangle = 2Dt.
\end{equation}
The average time to travel a distance $\sqrt{N}$ is then
\begin{equation}
\tau = {N\over2D} = \mbox{$\frac32$} N^3.
\end{equation}
Equation~\eref{moves} now tells us that one correlation time is, in the
worst case $E\to0$ limit, equivalent to $3N^4$ Monte Carlo moves.  This
gives a very steep increase in the CPU time required to perform the
simulation as a function of system size.  Only with recent advances in
parallel computation, as well as sophisticated multispin coding techniques,
has it become possible to perform these calculations with sufficient
accuracy to reveal the scaling behavior of the model.  Using the SP--1
supercomputer at Cornell University, Barkema, Marko and Widom~\cite{BMW94}
performed a calculation in which they measured the zero-field diffusion
coefficient for systems of a variety of lengths up to $N=100$.  They found
that the numerical data were well fit by the formula
\begin{equation}
N^2D=\frac{1}{3} \left( 1+5N^{-2/3} \right). 
\end{equation}
It was also found that for sufficiently large $N$, the drift velocity $v$
in a static field of strength $E$ was a function only of the scaling
variable $NE$.  The drift velocity was obtained for various values of $N$
and $NE$ up to $N=400$ and $NE=20$.  In the limit $N\to\infty$ and $E\to0$
it was found that the drift velocity was well fit by:
\begin{equation}
N^2v = NE \left[ \left( \frac{1}{3} \right)^2
     + \left( {2NE\over5} \right)^2 \right]^{1/2}
\end{equation}
for all values of $NE$.  Thus, $v$ has a finite, non-zero limit as
$N\to\infty$ at fixed $E$ and this limit is proportional to $E|E|$.  This
result is in agreement with the conclusions of Duke, Semenov and Viovy
\cite{DSV92} for a different but related model.

\section{The repton model as a particle model}
\label{secparticles}
\begin{figure}
\begin{center}
\psfigure{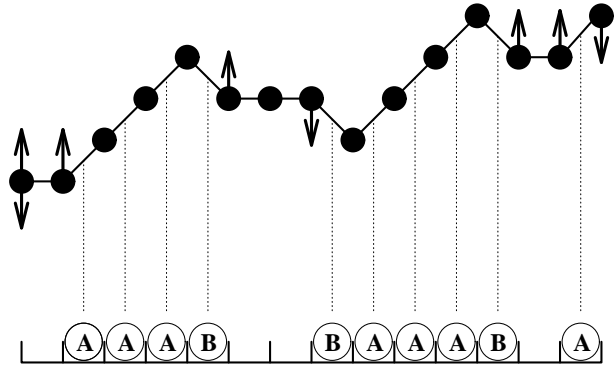}{8cm}
\end{center}
\caption{An illustration of the link between a configuration of the projected
repton model, and a configuration in the model with two species of
particles.}
\label{particles}
\end{figure}

Another representation of the repton model takes the form of a number of
particles with hard-sphere repulsion moving around on a one-dimensional
lattice.  In this representation each link between adjacent reptons in our
chain corresponds to a site on the new lattice, and we populate these sites
with two species of particles, which we will call types~A and~B,
corresponding to links which slope one way or the other in the projected
repton model.  For pairs of adjacent reptons which have the same
$x$-coordinate we leave the corresponding site on the new lattice empty.
This mapping is illustrated in Figure~\ref{particles} for one particular
configuration of the polymer.  The rules for the repton model are easily
expressed in the language of these particles:
\begin{enumerate}
\item In the interior of the chain the particles are conserved: they can
  move from site to site on the lattice but they are not created or
  annihilated.
\item A particle can only hop to a neighbouring site if that site is not
  already occupied by another particle.\footnote{The particles are a bit
    like Fermions in this respect.  Notice however that if they were true
    Fermions, particles of type~A and type~B would be able to coexist on
    one lattice site, because they are distinguishable.  Since our model is
  also a completely classical one, it is probably better simply to think of
  the particles as having hard-sphere interactions.}
\item A particle on one of the two end sites of the chain can leave the
  chain---it simply vanishes from the lattice.
\item If the end site of the chain is empty, a particle of either type can
  appear there.
\end{enumerate}

In the absence of an electric field, the rates for all particle moves are
equal.  On average, each particle on the chain tries to hop both to the
left and to the right once per unit of simulated time, particles of type~A
try to enter the chain at both the left and right ends once per unit of
time, and so do particles of type~B.  If an electric field is introduced,
type~A particles will hop to the left at the rate of $\exp(E/2)$ moves per
unit time and to the right at the rate of $\exp(-E/2)$.  Conversely, type~B
particles will have rates of $\exp(-E/2)$ and $\exp(E/2)$ moves per unit
time for hopping to the left and right.  Thus, an electric field draws
A-particles to the left, and B-particles to the right.

This new physical picture of the repton model is useful, since some things
which are not obvious in our other pictures are obvious with this one (and
{\it vice versa}).  For example, given that the electric field will push
type~A particles one way and type~B ones the other way, it is intuitively
reasonable if we start off with a mixture of the two types of particles,
that after a while one or more traffic jams are going to build up on the
chain, since the two particle types are trying to go in opposite directions
but cannot pass one another.  Furthermore, these traffic jams will be
unstable to fluctuations.  If one side of the jam becomes larger than the
other---if there are more particles of type~A for example than there are of
type~B---then the larger half will `push' the smaller half back, and,
ultimately, off the end of the chain.  When this happens, we end up with a
configuration of the chain dominated by one type of particle or the other.
What does this situation correspond to if we project it back onto the
original repton model?  It turns out that it corresponds to a polymer that
has got itself lined up along the electric field, so that the reptation
motion along the line of the molecule is in exactly the direction the field
is trying to push it.  This is a situation that is frequently observed in
electrophoresis experiments.

Movement of the entire chain along the $x$-axis corresponds to the
propagation of particles all the way across the lattice from one end of the
system to the other.  For example, the centre of mass of the chain will
move one step in the positive $x$ direction every time a particle of type~A
enters the system at the left-hand end and leaves it at the right.  It will
do the same if a particle of type~B moves from right to left.  Particles of
these types going in the opposite directions will move the chain one lattice
spacing in the negative $x$ direction.

\begin{figure}
\begin{center}
\psfigure{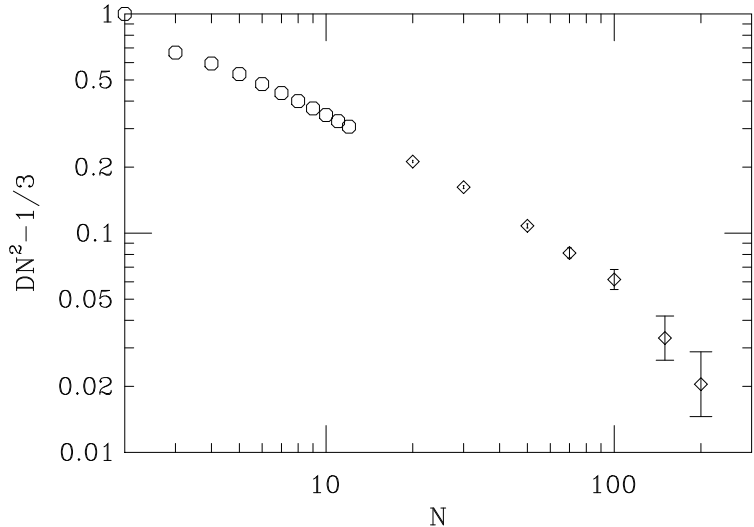}{10cm}
\end{center}
\caption{Log-log plot of $DN^2-1/3$ as a function of $N$. In the range
  $20<N \leq 100$, the data is well described by $DN^2=1/3+5/3 N^{-2/3}$,
  as was reported by Barkema, Marko and Widom\protect\cite{BMW94}.  The new
  data for $N \ge 100$ suggests that maybe for larger $N$ this fit is not
  appropriate any more.}
\label{Dplot}
\end{figure}

Let us denote by $n_{A\ra}$ and $n_{B\ra}$ the number of particles of
types~A and~B which cross the system from left to right in a certain time,
and by $n_{A\la}$ and $n_{B\la}$ the number which go the other way.  In the
long-time limit the displacement $\Delta x$ of the chain is then given by
\begin{equation}
\Delta x = (n_{A\ra} + n_{B\la}) - (n_{A\la} + n_{B\ra}).
\end{equation}
In the $E=0$ case, the two types of particles diffuse in exactly the same
fashion, and are therefore indistinguishable, a fact which we can exploit
to improve the accuracy of our simulations as follows.  If we perform a
simulation of the particle model with $E=0$, the movements of the particles
will be the same regardless of their type, which means that there is no
need to know what types they actually are.  All we need to do is keep track
of the number of particles crossing the system in each direction during the
course of the run and then we can assign types~A and~B to these afterwards
in any fashion we choose.  In fact, the best statistics are to be obtained
by considering every possible such assignment, and averaging over them all.
The mean square displacement of the system $\langle (\Delta x)^2 \rangle$
is then given by\begin{eqnarray} \langle (\Delta x)^2 \rangle & = & \langle
  [(n_{A\ra} - n_{B\ra})
  - (n_{A\la} - n_{B\la})]^2 \rangle \nonumber\\
  & = & \langle (n_{A\ra} - n_{B\ra})^2 \rangle +
  \langle (n_{A\la} - n_{B\la} )^2 \rangle \nonumber\\
  & = & (n_{A\ra} + n_{B\ra}) + (n_{A\la} + n_{B\la})\nonumber\\
  & = & n_\ra + n_\la.
\label{cross}
\end{eqnarray}
The second line here follows since the assignment of As and Bs is
uncorrelated with the directions in which the particles cross the system
and the third follows from properties of random walks in one dimension.
Thus, all we have to do to obtain the diffusion coefficient in a simulation
is count the total numbers of particles $n_\ra$ and $n_\la$ which cross the
lattice in either direction.  The implicit average over particle types
in~\eref{cross} gives us an improvement in the efficiency of the
calculation of at least an order of magnitude for the typical system sizes
we are studying here.  We give some preliminary results obtained using this
method in Table~1 and Figure~\ref{Dplot}.

\begin{table}
\begin{center}
\begin{tabular}{ccc}
$N$ & $DN^2-\frac13$ \\
\hline
\hline
20  & $0.2123\pm0.0027$ \\
30  & $0.1622\pm0.0025$ \\
50  & $0.1079\pm0.0023$ \\
70  & $0.0812\pm0.0044$ \\
100 & $0.0615\pm0.0065$ \\
150 & $0.0332\pm0.0077$ \\
200 & $0.0205\pm0.0070$ \\
\hline
\hline
\end{tabular}
\end{center}
\medskip
\caption{The quantity $DN^2-\frac13$ measures the deviation of the
  diffusion constant $D$ from its infinite $N$ limit of
  $1/(3N^2)$~\protect\cite{Prahofer94}.}
\end{table}

\subsection{Particle densities}
\label{densities}
In order to calculate the numbers $n_\ra$ and $n_\la$ appearing in
Equation~\eref{cross}, we need to keep a record for each particle of
whether it entered from the left or the right end of the lattice.  Let us
label each particle with an L or an R, depending on the end at which it
entered.  It turns out that the average densities of L-type and R-type
particles possess some interesting features, suggesting a possible route to
an exact solution of the repton model.

\begin{figure}
\begin{center}
\psfigure{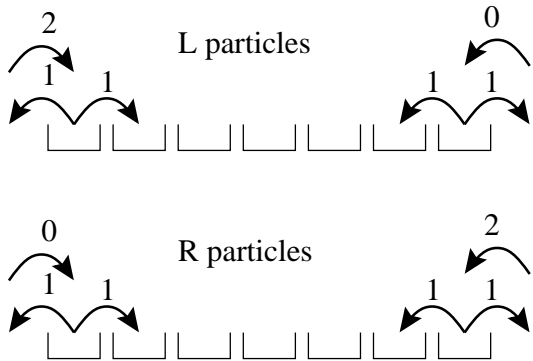}{7cm}
\end{center}
\caption{Illustraction of the dynamics of the L and R particles described
  in the text.}
\label{LR}
\end{figure}

The dynamics of the L and R particles is very similar to that of the A and
B particles we considered in the last section (they exclude each other,
they try to hop left and right with rate 1, they disappear from either end
with rate 1), except that L particles enter only from the left, and
R particles from the right, both with rate 2 (whereas A and B particles
entered from either side with rate 1).  These rules are depicted in
Figure~\ref{LR}.

The total particle density at any site is still equal to $\frac23$, just as
before.  However, since L particles do not enter at the right end, R
particles do not enter at the left, and the two types are unable to pass
one another, the system divides into two domains, each containing just one
type of particle as well as vacancies, separated in the middle by an
interface.  The position of this interface can fluctuate and at times may
move all the way to one end of the lattice, allowing particles of one type
or the other to pass through the system causing a shift in the centre of
mass of the repton chain.  However, it is clear that on average the density
of particles of type~L will be greater at the left-hand end of the system
and {\it vice versa\/} for particles of type~R.  Because of symmetry
between L and R particles, the average density of R particles at a site $i$
is equal to the density of L particles at site $N-i$:
\begin{equation}
\rho_R(i) = \rho_L(N-i) \equiv \rho(i).
\end{equation}

We have measured the density profile $\rho(i)$ using our multispin Monte
Carlo method for systems of size $N=20$, 50, 100 and 200.  For each value
of $N$ we have performed three runs, equivalent, through the magic of
multispin coding, to 192 individual simulations, since we performed the
runs on a 64-bit computer.  The runs were done over $5\times10^8$ moves for
$N=20$ and 50, and $5\times10^9$ moves for $N=100$ and 200.  The
simulations were started with random configurations with an average
particle density of $\frac23$ (the expected equilibrium value) and random
choices for the interface position.  To ensure thermalization, we initially
ran the simulations for an additional 10\% of the total number of moves and
discarded the data.

\begin{figure}
\begin{center}
\psfigure{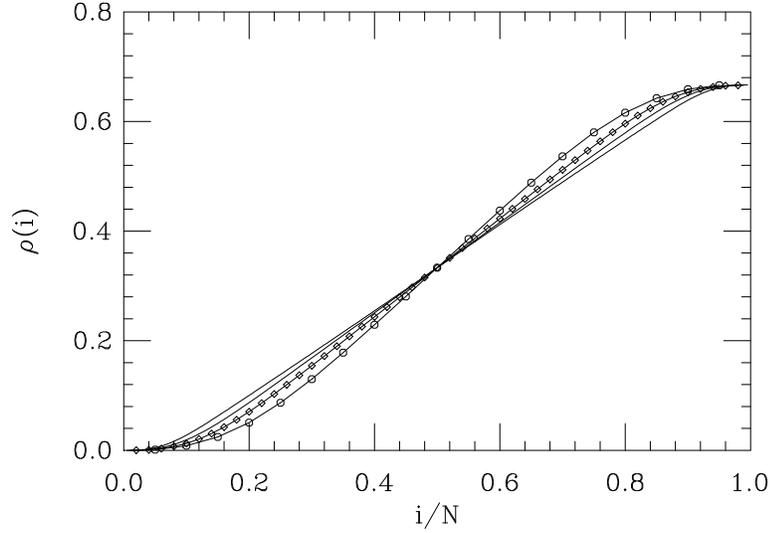}{10cm}
\end{center}
\caption{Density of R particles at site $i$, as a function of
  distance along the chain (as a fraction of the total chain length.}
\label{linfig}
\end{figure}

\begin{figure}
\begin{center}
  \psfigure{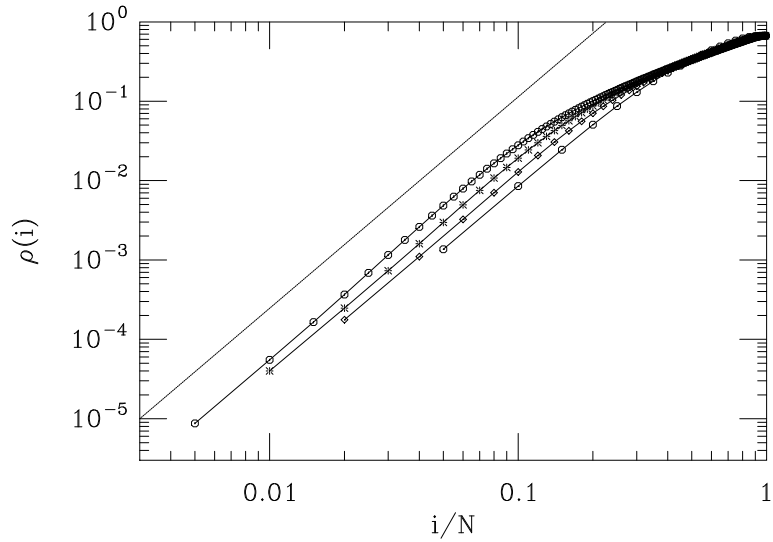}{10cm}
\end{center}
\caption{Log-log plot of the density of R particles at site $i$, as a
  function of distance along the chain.  For low values of $i$, the density
  appears to scale as a power-law.  The dashed line is added a guide to the
  eye; it is a power-law with exponent $\frac83$.}
\label{loglogfig}
\end{figure}

In Figures~\ref{linfig} and~\ref{loglogfig} we have plotted $\rho(i)$ as a
function of $i$ on linear and logarithmic scales.  In the latter, we
observe power-law scaling of $\rho(i)$ as a function of $i$ for small $i$:
\begin{equation}
\rho_R(i) \sim i^{\alpha},
\end{equation}
with an exponent which is independent of the length of the chain:
\begin{equation}
\alpha = 2.70 \pm 0.04.
\end{equation}
We conjecture that the actual value of this exponent may be $N^{-8/3}$ and
that it may be related to the sub-dominant scaling of the diffusion
coefficient observed by Barkema, Marko and Widom~\cite{BMW94}.

It is in fact possible to compute the value of the diffusion coefficient
from a knowledge of the density profile: the rate with which L particles (R
particles) leave the system at the right-hand (left-hand) end, is strictly
equal to their density at the last site.  Using Equation~\eref{cross}, we
thus obtain:
\begin{equation}
D = \half\frac{\partial}{\partial t} \langle (\Delta x)^2 \rangle = \rho(1).
\label{diffusion}
\end{equation}

This formalism suggests a possible route to an exact solution of the repton
model.  For a class of models known as asymmetric exclusion models,
Derrida~\etal~\cite{DEHP93} have derived an analytic expression for the
particle density profile for any $N$.  The dynamics of these models is
closely related to that of the particle version of the repton model
presented here, and it is possible that the result might be extended to the
repton model as well.  If so, Equation~\eref{diffusion} would then tell us
the exact value of the diffusion constant for any $N$.

\section{Conclusions}
\label{conclusions}
We have described the repton model of the agarose gel electrophoresis of
DNA, as well as its projection onto a one-dimensional reptation model and a
one-dimensional particle model.  We have reviewed exact analytic results on
the model for short chain lengths and numerical studies of longer ones.  In
addition we have described a new algorithm for simulating the model using
the particle representation and presented preliminary results from
simulations using this algorithm.  In closing we have pointed out that the
diffusion constant of the  polymer is related in a simple way to the
density profile of the particles which suggests a possible route to an
exact solution of the model.

\section{Acknowledgements}
One of us (GTB) would like to thank the Santa Fe Institute for their
hospitality whilst this work was carried out.  This research was funded in
part by the DOE under grant number DE--FG02--9OER40542 and by the Santa Fe
Institute and DARPA under grant number ONR N00014--95--1--0975.


\begin{thebibliography}{99}
\bibitem{deGennes71}
{\frenchspacing P. G. de Gennes, J. Chem. Phys. {\bf55}, 572 (1971).}
\bibitem{Rubinstein87}
{\frenchspacing M. Rubinstein, Phys. Rev. Lett. {\bf59}, 1946 (1987).}
\bibitem{Duke89}
{\frenchspacing T. A. J. Duke, Phys. Rev. Lett. {\bf62}, 2877 (1989).}
\bibitem{Duke90a}
{\frenchspacing T. A. J. Duke, J. Chem. Phys. {\bf93}, 9049 (1990).}
\bibitem{Duke90b}
{\frenchspacing T. A. J. Duke, J. Chem. Phys. {\bf93}, 9055 (1990).}
\bibitem{LK92}
{\frenchspacing J. M. J. van Leeuwen and A. Kooiman, Physica A {\bf184}, 79
  (1992).}
\bibitem{KL93a}
{\frenchspacing A. Kooiman and J. M. J. van Leeuwen, Physica A {\bf194},
  163 (1993).}
\bibitem{KL93b}
{\frenchspacing A. Kooiman and J. M. J. van Leeuwen,
  J. Chem. Phys. {\bf99}, 2247 (1993).}
\bibitem{WVD91}
{\frenchspacing B. Widom, J. L. Viovy and A. D. Defontaines, J. Phys. I
  France {\bf1}, 1759 (1991).}
\bibitem{Widom96}
{\frenchspacing B. Widom, Physica A, in press (1996).}
\bibitem{BMW94}
{\frenchspacing G. T. Barkema, J. F. Marko and B. Widom, Phys. Rev. E
  {\bf49}, 5303 (1994).}
\bibitem{BCM96}
{\frenchspacing G. T. Barkema, C. Caron and J. F. Marko, Biopolymers
  {\bf38}, 665 (1996).}
\bibitem{Prahofer94}
{\frenchspacing M. Pr\"ahofer, Diplomarbeit, Ludwig-Maximilians-Universit\"at
  M\"unchen, (1994).}
\bibitem{DSV92}
{\frenchspacing T. A. J. Duke, A. N. Semenov, and J. L. Viovy, Phys. Rev.
  Lett. {\bf 69}, 3260 (1992).}
\bibitem{DEHP93}
{\frenchspacing B. Derrida, M. R. Evans, V. Hakim, and V. Pasquier,
J. Phys. A Math. Gen. {\bf 26}, 1493 (1993).}
\end{thebibliography}
\end{document}